\documentclass[journal=apcach,manuscript=article]{achemso}
\usepackage{placeins}
\usepackage[version=3]{mhchem} 
\usepackage{units}

\usepackage{xcolor}
\usepackage{graphicx}
\newcommand{\fixme}[1]{{\color{black}{#1}}} 
\renewcommand{\vec}[1]{\boldsymbol{#1}}   
\newcommand{\mat}[1]{\mathbf{#1}}         
\author{Torsha Moitra}
\email{torsha@iitbhilai.ac.in}
\affiliation{Department of Chemistry, Indian Institute of Technology Bhilai, Durg, Chhattisgarh 491001, India}
\alsoaffiliation{Department of Physical and Theoretical Chemistry, Faculty of Natural Sciences, Comenius University, 84215 Bratislava, Slovakia}
\author{Lukas Konecny}
\affiliation{Institute of Inorganic Chemistry, Slovak Academy of Sciences, 84536 Bratislava, Slovakia}
\alsoaffiliation{Department of Inorganic Chemistry, Faculty of Natural Sciences, Comenius University,
84215 Bratislava, Slovakia}
\alsoaffiliation{Hylleraas Centre for Quantum Molecular Sciences, Department of Chemistry, UiT The Arctic University of Norway, 9037 Tromsø, Norway}
\author{Michal Repisky}
\affiliation{Hylleraas Centre for Quantum Molecular Sciences, Department of Chemistry, UiT The Arctic University of Norway, 9037 Tromsø, Norway}
\alsoaffiliation{Department of Physical and Theoretical Chemistry, Faculty of Natural Sciences, Comenius University, 84215 Bratislava, Slovakia}
\title[]
  {Attosecond Charge Migration Induced by Core-Level Ionization: A Relativistic Real-Time Time-Dependent Density Functional Theory Perspective}


\begin{document}

%
%

\begin{abstract}
Core-ionization induced charge migration in molecular systems has been widely studied, but the influence of relativistic effects on such dynamics remains relatively unexplored from a relativistic theory perspective. This issue is particularly important because core orbitals exhibit strong scalar-relativistic and spin-orbit imprints, and these effects are further enhanced in heavy-element systems, where relativistic signatures also alter the valence orbitals. To address this, we formulate four-component Dirac-Coulomb and two-component atomic mean-field exact two-component Hamiltonian approaches for studying charge migration within real-time time-dependent density functional theory. The time-dependent induced electric dipole moment is used as a marker of the characteristic charge-migration time scales. We apply the method to nitrosobenzene (N $1s$ electron removal), iodoacetylene (I $2p$ electron removal) and interhalogen series (I $2p$, Cl $2p$ electron removal). For iodoacetylene, the charge migration timescale is relatively insensitive to the choice of Hamiltonian but they show weak relativistic phase and amplitude modifications. However, it would be misleading to conclude that relativistic effects are generally unimportant. In interhalogen systems, we observe significant deviations between relativistic and non-relativistic dynamics. These results demonstrate that relativistic signatures in core-ionization-induced charge migration are not determined by heavy-atom core ionization alone, but by the relativistic sensitivity of the valence orbital manifold that drives core-hole screening and charge redistribution. Moreover, consistently across all systems, we observe a delay in the emergence of the relativistic imprints.
\end{abstract}

\section{Introduction}
Charge migration is one of the most widely studied processes in attosecond molecular dynamics.~\cite{Worner2017} It refers broadly to the coherent redistribution of electronic charge driven by a nonstationary electronic wavepacket and occurring on a time scale faster than nuclear motion. In many experimentally relevant situations, such wavepackets are generated by ultrafast ionization, leading to coherent hole dynamics. This purely electronic motion was first predicted by Cederbaum and Zobeley~\cite{Cederbaum1999} in the late 1990s and has since become a central concept in attochemistry. Experimentally, charge migration and related ultrafast hole dynamics have been investigated in a range of molecular systems, including phenylalanine~\cite{Belshaw2012,Calegari2014}, iodoacetylene~\cite{Kraus2015}, tryptophan~\cite{Trabattoni2019}, silane~\cite{Matselyukh2022}, and other small- and medium-sized molecules.\cite{Calegari2016, Maansson2021, Driver2026} These studies have established that charge migration is highly sensitive to the electronic structure of the ionized molecule, particularly to the nature of the valence orbitals that participate in the ensuing charge dynamics.

Real-time electronic-structure methods~\cite{Goings2018,Li2020,Ofstad2023,Kadek2024,moitra2026} provide a direct theoretical route for following such dynamics. Among them, real-time time-dependent density functional theory (RT-TDDFT)~\cite{Theilhaber1992, Yabana1996, Lopata2011, Ullrich2011, Repisky2015, Konecny2016, Goings2016} has become a practical approach because it allows the time evolution of the electron density to be monitored in molecular systems at moderate computational cost, making it a popular method of choice for simulating \fixme{core-level spectroscopies~\cite{Kadek2015,Kasper2018,Ye2022,Ye2024,Ye2026} or} dynamical processes in molecules including charge migration.~\cite{Petrone2014, Bruner2017, Folorunso2021, Khalili2021, Mauger2022, Sinha2026} However, charge migration is a complex process: the dynamics depend on the preparation of the initial ionized state, the description of electron correlation, self-interaction error, exchange-correlation memory effects, and role of Auger--Meitner decay.~\cite{Burhop1972, Drescher2002, Smirnova2009, Kuleff2010, Wahyutama2026} The initially populated cationic wavepacket is often difficult to define uniquely because it depends on pulse parameters.~\cite{Kuleff2014} A common approach for simplifying this dependence in simulations is provided by sudden ionization approximation~\cite{Pickup1977, Hedin2002} as it represents a more controlled alternative whereby an electron is removed instantaneously from a selected orbital, and the resulting nonstationary density is propagated in time. In the case of ionization from core orbitals which are spatially localized and energetically separated from the valence manifold, the initial core-hole state is more clearly defined than many valence-ionized wavepackets.~\cite{Kuleff2016,Bruner2017,Hua2025}

Attosecond charge migration induced by core ionization is particularly interesting since it is a process driven purely by electron correlation before the onset of nuclear motion.~\cite{Kuleff2016} The localized core hole perturbs the valence density which undergoes charge redistribution. Consequently, the early-time charge dynamics can be viewed as the coherent valence response to a localized core perturbation~\cite{Kuleff2016}, whereas longer-time dynamics may be modified or quenched by core-hole decay. The core hole has a finite lifetime due to Auger--Meitner decay~\cite{Burhop1972}, subsequent multihole formation or related electronic relaxation processes.~\cite{Picon2018,Hua2025} For light-element K-edge holes, the core hole decay often occurs on a few-femtosecond time scale, while deeper or heavier-element core holes may exhibit even more pronounced lifetime effects.~\cite{Krause1979} The relevant charge-migration window is therefore the sub-femtosecond to few-femtosecond regime in which valence screening develops before nuclear motion and major electronic decay processes dominate.

Core-ionization-induced charge migration dynamics, particularly in systems containing heavier elements, may be strongly influenced by relativistic effects. However, most theoretical studies of these phenomena to date have employed nonrelativistic electronic-structure methods. In our opinion, this represents a significant limitation, since core orbitals are particularly sensitive to relativity: scalar-relativistic effects alter their energies and radial distributions, while spin--orbit coupling modifies their spatial structures.~\cite{Dyall_book} Moreover, relativistic effects propagate into the valence orbital manifold~\cite{Pyykko1988}, with potentially direct consequences for charge migration dynamics. Therefore, \emph{relativistic} electronic-structure theory is required not only for an accurate description of core levels but also for capturing the response in the valence shells.

The main goal of this work is to examine the effects of relativity on core-hole-induced charge migration using relativistic RT-TDDFT. We employ both the four-component (4c) Dirac--Coulomb Hamiltonian~\cite{Repisky2015} and the atomic mean-field exact two-component (amfX2C) Hamiltonian~\cite{Knecht2022,Moitra2023} to describe relativistic electron dynamics and compare the results with nonrelativistic reference calculations. The induced electric dipole moment serves as a compact indicator of charge-migration time scales, while the time-dependent induced charge density provides a real-space picture of the electronic response.
\fixme{However, the present single-determinant, adiabatic RT-TDDFT approach cannot explicitly describe correlated shake-up (2h1p) satellite states or Auger--Meitner decay,
which require more sophisticated many-electron treatments, as pioneered for stationary states in Ref.~\citenum{Hillier1976,Cederbaum1980,Chaudhuri1999}.
}
We first validate the implementation on nitrosobenzene and then investigate relativistic effects on charge migration in iodoacetylene and interhalogen compounds. In these systems, substantial deviations between relativistic and nonrelativistic dynamics are observed when the valence orbitals governing the dynamics are split by spin--orbit coupling.

\section{Theory}

The implemented methodology for the theoretical study of ultrafast charge migration is based on relativistic real-time time-dependent density functional theory (RT-TDDFT)~\cite{Repisky2015,Konecny2016} as implemented in our ReSpect program~\cite{Repisky2020,Repisky2025} and employs the sudden-ionization approximation,~\cite{Pickup1977, Hedin2002} in which an electron is instantaneously removed from an $N$-electron neutral molecule in its ground state. The resulting cationic system is in a nonstationary state that evolves in time, giving rise to electronic charge migration. Within the RT-TDDFT framework, the initial nonstationary cationic state is characterized by the diagonal one-particle reduced density matrix
\begin{equation}
    \mat{D}^{\text{MO}}_{+}
    =
    \begin{pmatrix}
        \mat{I}_{\text{oo}}^{+} & \mat{0}_{\text{ov}} \\
        \mat{0}_{\text{vo}}     & \mat{0}_{\text{vv}}
    \end{pmatrix}
    ;\qquad
    \mat{I}_{\text{oo}}^{+} = \text{diag}\big(1,...,1,0,1,...,1\big)
\end{equation}
where we partition the molecular orbital (MO) space into occupied (o) and virtual (v) subspaces.
$\mat{I}_{\text{oo}}^{+}$ denotes the diagonal $N\!\times\!N$ matrix with unit diagonal elements, except for a single zero representing the hole created in a particular occupied MO. The remaining blocks of $\mat{D}^{\text{MO}}_{+}$ are zero matrices, implying that, within the sudden-ionization picture, initially all virtual MOs are vacant and there are no coherences between occupied and virtual MOs. In the relativistic four-component theory, the virtual space also includes all negative-energy eigenstates of the Dirac--Fock operator.

Within our RT-TDDFT framework, the density matrix of the nonstationary cationic state evolves according to the Liouville--von Neumann equation of motion
\begin{equation}
        i\frac{\partial \mathbf{D}^{\text{MO}}_{+}(t)}{\partial t}
        =
        \left[ \mathbf{F}^{\text{MO}}(t), \mathbf{D}^{\text{MO}}_{+}(t) \right]
\label{eq:vonNeumann}
\end{equation}
Here, $t$ is the time variable, $i$ is the imaginary unit, and $[\mat{F}^{\text{MO}},\mat{D}^{\text{MO}}_{+}]$ denotes the commutator of the density matrix and the Fock matrix, with the latter driving the time evolution of the cationic system. Both matrices are represented in the ground-state SCF MO basis, with MOs expanded in Gaussian-type atomic orbitals (AOs), which provide a compact description of core states and core holes and enable the efficient evaluation of exact exchange in hybrid DFT.
The Liouville--von Neumann equation~\eqref{eq:vonNeumann} is solved using the second-order Magnus (exponential midpoint) propagator~\cite{Magnus1954},
\begin{equation}
    \mat{U}(t+\Delta t,t)
    \approx
    \exp\left[-i\mat{F}^{\text{MO}}\left(t+\frac{\Delta t}{2}\right) \Delta t\right]
    \label{eq:magnusMidpoint}
\end{equation}
which propagates the density matrix according to
\begin{equation}
    \mat{D}^{\text{MO}}_{+}(t+\Delta t)
    =
    \mat{U}(t+\Delta t, t)\, \mat{D}^{\text{MO}}_{+}(t)\, \mat{U}^\dagger (t+\Delta t, t)
\label{eq:evolDMatMagnus}
\end{equation}
However, the dependence on the future-time Fock matrix $\mat{F}^{\text{MO}}\left(t+\tfrac{\Delta t}{2}\right)$ in Eq.~\eqref{eq:magnusMidpoint} requires a sophisticated iterative extrapolation--interpolation scheme to ensure numerical stability of the time propagation. Details of the implemented solver are discussed in Ref.~\citenum{Repisky2015}.

At each time step, the propagated density and Fock matrices are transformed between the MO and AO representations using the ground-state MO coefficient matrix $\mat{C}$:
\begin{align}
        \mat{D}^{\text{AO}}_{+}(t)
        \,=\,
        \mat{C}\mat{D}^{\text{MO}}_{+}(t)\mat{C}^{\dagger}
        ;\qquad\qquad
        \mathbf{F}^{\text{MO}}(t)
        \,=\,
        \mathbf{C}^{\dagger}\mathbf{F}^{\text{AO}}(t)\mathbf{C}
\label{eq:aoDensity}
\end{align}
Our relativistic implementation employs an integral-direct algorithm for evaluating the time-dependent AO Fock matrix and supports both the four-component (4c)~\cite{Repisky2015} and atomic mean-field exact two-component (amfX2C)~\cite{Konecny2016,Knecht2022} Hamiltonian frameworks, with $\mat{D}^{\text{AO}}_{+}=\mat{D}^{\text{4c}}_{+}$ and $\mat{D}^{\text{AO}}_{+}=\mat{D}^{\text{2c}}_{+}$, respectively:
\begin{equation}
    \mathbf{F}^{\text{AO}}(t)
    =
    \begin{cases}
        \mathbf{F}^{\text{4c}}(t)
        =
        \mathbf{h}^{\text{4c}}
        +
        \mathbf{G}^{\text{4c}}[\mathbf{D}^{\text{4c}}_{+}(t)]
        +
        \mathbf{V}^{\text{4c}}_{\text{xc}}[\vec{\rho}^{\text{4c}}_{+}(t),\vec{\nabla}\vec{\rho}^{\text{4c}}_{+}(t)]
    \\[0.4cm]
        \mat{F}^{\text{amfX2C}}(t)
        =
        \mat{\tilde{h}}^{\text{2c}}
        +
        \mat{G}^{\text{2c}}[\mat{D}^{\text{2c}}_{+}(t)]
        +
        \Delta\mat{G}^{\text{2c}}_{\bigoplus}
        +
        \mathbf{V}^{\text{2c}}_{\text{xc}}[\vec{\rho}^{\text{2c}}_{+}(t),\vec{\nabla}\vec{\rho}^{\text{2c}}_{+}(t)]
        +
        \Delta\mat{V}^{\text{2c,xc}}_{\bigoplus}
    \end{cases}
\end{equation}
The matrices $\mat{h}$, $\mat{G}$, and $\mat{V}_{\text{xc}}$ denote the one-electron Hamiltonian, the two-electron Coulomb and exact-exchange contribution evaluated from the corresponding time-dependent AO density matrix $\mat{D}_{+}(t)$, and the exchange--correlation (XC) potential evaluated from the corresponding time-dependent electron density vector $\vec{\rho}_{+}(t)$ and its gradient $\vec{\nabla}\vec{\rho}_{+}(t)$, respectively. In relativistic DFT, the electron density vector comprises the electron number density and the three Cartesian components of the electron spin density. The spin density is treated using the noncollinear ansatz, which preserves the rotational invariance of nonrelativistic spin-density functionals within the relativistic framework~\cite{VanWuellen2002}. In the present work, we employ the noncollinear approach proposed by Scalmani and Frisch~\cite{Scalmani2012}, as detailed in Refs.~\citenum{Komorovsky2019,Repisky2020}. All 4c Fock matrix elements are expressed in an AO basis satisfying the restricted kinetic balance (RKB) condition between its large and small components~\cite{Stanton1984}. In the amfX2C formulation, the tilde denotes quantities subjected to the X2C (picture-change) transformation to the two-component (2c) representation. Accordingly, $\mat{\tilde{h}}^{\text{2c}} = \Big[\mat{U}^{\dagger}\mat{h}^{\text{4c}}\mat{U}\Big]^{++}$, where $\mat{U}$ is the unitary X2C decoupling matrix~\cite{Kutzelnigg2005,Liu2007,Ilias2007} and $[\ldots]^{++}$ denotes the positive-energy block retained after the X2C transformation. The two-electron and XC contributions are evaluated in the untransformed 2c AO basis, as indicated by the absence of a tilde. The corresponding picture-change corrections are included approximately through the atomic mean-field terms $\Delta\mat{G}^{\text{2c}}_{\bigoplus}$ and $\Delta\mat{V}^{\text{2c,xc}}_{\bigoplus}$ for the two-electron and XC contributions~\cite{Knecht2022}, respectively. For further details on the theoretical formulation and implementation of the 4c and amfX2C approaches in our program, we refer the reader to Refs.~\citenum{Repisky2020,Repisky2025} and the references therein.

In agreement with previous works, the resulting dynamics were interpreted in terms of the electron hole density, which measures the local change in electron density of the cation relative to the neutral ground state~\cite{Cederbaum1999}
\begin{equation}
    h(\vec{r},t) = \rho_{0}(\vec{r}) - \rho_{+}(\vec{r},t)
\end{equation}
where $\rho_{0}$ is the ground-state electron density of the neutral molecule and $\rho_{+}$ is the time-dependent electron density of the cation. Positive values of the hole density (hole accumulation) correspond to electron depletion, whereas negative values indicate electron accumulation relative to the neutral ground state.

In our relativistic implementation, the hole density is evaluated in AOs as the trace of the product of the charge distribution matrix $\mat{\Omega}$ and the hole density matrix:
\begin{equation}
    h(\vec{r},t)
    =
    \begin{cases}
        h^{\text{4c}}(\vec{r},t)
        =
        \text{Tr}\Big[ \mat{\Omega}^{\text{4c}}(\vec{r})
                 \left( \mat{D}^{\text{4c}}_{0} - \mat{D}^{\text{4c}}_{+}(t) \right)
                 \Big]
        \\[0.4cm]
        h^{\text{2c}}(\vec{r},t)
        =
        \text{Tr}\Big[ \mat{\tilde{\Omega}}^{\text{2c}}(\vec{r})
                 \left( \mat{D}^{\text{2c}}_{0} - \mat{D}^{\text{2c}}_{+}(t) \right)
                 \Big]
    \end{cases}
    \label{eq:hole-density}
\end{equation}
The elements of the 4c charge distribution matrix are products of two 4c basis functions,
\begin{equation}
    \begin{aligned}
    \Omega^{\text{4c}}_{\mu\nu}(\vec{r})
    & =
    \Big(X^{\text{RKB}}_{\mu}(\vec{r})\Big)^{\dagger} X^{\text{RKB}}_{\nu}(\vec{r})
    \\[0.2cm]
    X_{\mu}^{\text{RKB}}(\vec{r})
    & =
    \begin{bmatrix}
        \text{I}_{2} & 0_{2}
        \\
        0_{2} & \frac{1}{2c}(\vec{\sigma}\cdot\vec{p})
    \end{bmatrix}
    f_{\mu}(\vec{r})
    \end{aligned}
\end{equation}
where the 4c basis functions $X_{\mu}^{\text{RKB}}$ obey the restricted kinetic balance (RKB) condition between their large and small components~\cite{Stanton1984}. Here, $f_{\mu}(\vec{r})$ denotes a normalized scalar AO basis function specified by the user, as in the nonrelativistic theory. Consistent with the discussion above, $\mat{\tilde{\Omega}}^{\text{2c}}$ equals to $\Big[\mat{U}^{\dagger}\mat{\Omega}^{\text{4c}}\mat{U}\Big]^{++}$ and denotes the X2C-transformed charge distribution matrix.

Finally, the hole density $h(\vec{r},t)$ can also be evaluated in the MO basis using a formula analogous to Eq.~\eqref{eq:hole-density}. Spatial integration of the hole density yields the time-dependent hole number, which can be further decomposed into molecular-orbital contributions~\cite{Breidbach2003}, referred to as hole occupation numbers or simply hole occupations. These occupations characterize the ionization-induced time-dependent population redistribution among the valence occupied and virtual orbitals and are calculated for the $p$th MO as:
\begin{equation}
   n^{\text{h}}_{p}(t) = \Big[ \mat{D}_{0}^{\text{MO}} - \mat{D}_{+}^{\text{MO}}(t) \Big]_{pp}
   \label{eq:hole-occupation-MO}
\end{equation}

\section{Computational Details}
The electron dynamics following the core-level ionization were simulated within the real-time time-dependent density functional theory (RT-TDDFT) at three Hamiltonian levels: non-relativistic (1c), atomic mean-field exact two-component amfX2C (2c)~\cite{Konecny2016,Knecht2022,Moitra2023} and four-component Dirac--Coulomb (4c)~\cite{Repisky2015} Hamiltonians, as implemented in the ReSpect software.~\cite{Repisky2020,Repisky2025}
\fixme{The initial core-hole state was generated by removing an electron from a selected ground-state orbital, creating a nonstationary cationic electronic wavepacket.
For $2p$ core ionization, an electron was removed from the $2p_z$ orbital in the nonrelativistic (1c) calculations, whereas in the relativistic (2c and 4c) calculations, an electron was removed from one of the spinors belonging to the $2p_{1/2}$ level.
The sensitivity of the charge migration dynamics to different initial ionization configurations within the $2p$ shell (L-edge) was also examined, revealing
negligible differences in the induced dipole dynamics, as shown in the Supporting Information. The population of the initially created core hole remained unchanged throughout the reported simulations.} The following field-free electron dynamics were obtained by propagating the one-electron density matrix, with a fixed nuclei. The time propagation was performed with a time-step of 0.01 au for a total time of \unit[2.5]{fs} (for nitrosobenzene), \unit[5]{fs} (for iodoacetylene) and \unit[3.5]{fs} (for interhalogen compounds).
\fixme{The iterative propagation employed a convergence threshold of $10^{-7}$ au, typically reached within 4--5 microiterations per time step.}
For all simulations, we use PBE0 exchange correlation functional~\cite{Slater1951, Perdew1996, Perdew1997Erratum, Adamo1999} with uncontracted aug-cc-pVDZ basis for C, H, N, O and aug-dyall-vDZ for halogens.~\cite{Kendall1992,Dyall-TCA-115-441-2006}
The geometries used for the simulation are provided in the Supplementary Information file.

\section{Results and Discussion}
In this section we examine how relativistic effects influence the electron dynamics following sudden core-level ionization. The discussion is organized around three sets of molecules: nitrosobenzene (Section~\ref{sec:nitrosobenzene}), iodoacetylene (Section~\ref{sec:iodoacetylene}) and interhalogen compounds (Section~\ref{sec:interhalogen}). First we consider nitrogen $1s$-ionized nitrosobenzene as a benchmark for the charge-migration dynamics and establish the time-dependent induced electric dipole moment as a compact descriptor of the dominant charge redistribution process. We then investigate iodine $2p$-ionized iodoacetylene, where comparison of the nonrelativistic (1c) and relativistic (2c and 4c) dynamics provides two key insights: (i) identify the delayed emergence of relativistic imprints and (ii) provide orbital-level analysis for the same. Finally, we study the interhalogen series, IF, ICl, IBr and IAt to assess how the relativistic response varies with the electronic structure of the participating valence molecular orbitals.

\subsection{Nitrosobenzene}\label{sec:nitrosobenzene}
\begin{figure}[htb!]
    \centering
    \includegraphics[width=1.0\linewidth]{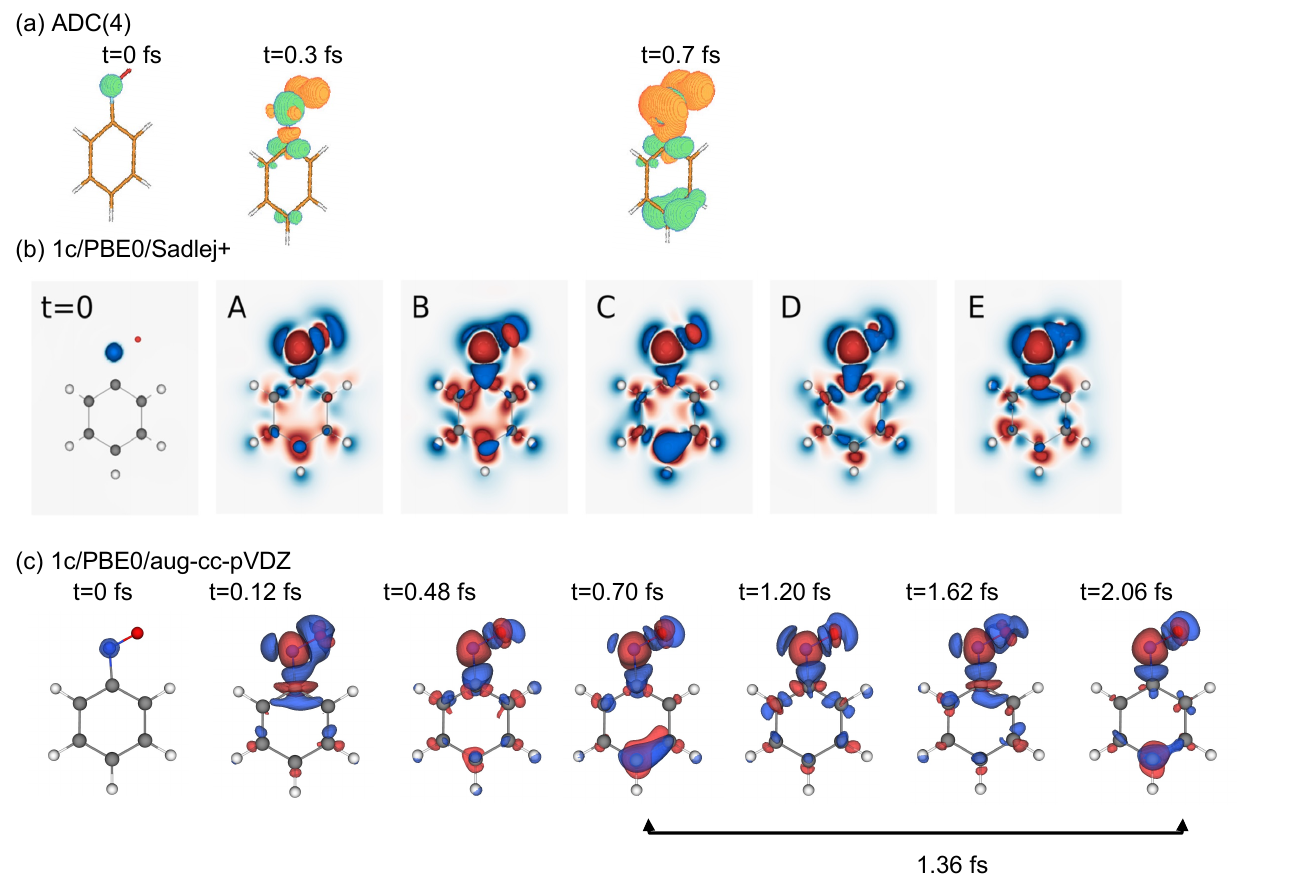}
    \caption{Nitrosobenzene: Time evolution of the hole density following sudden ionization from the N $1s$ orbital, as defined in Eq.~\eqref{eq:hole-density}. Charge migration computed at (a) ADC(4) level and (b) 1c/PBE/Sadlej+ level as reported in Refs.~\citenum{Kuleff2016} and~\citenum{Bruner2017}, respectively. (c) Our simulations are performed at 1c/PBE0/aug-cc-pVDZ level, and the hole densities are plotted at iso-value 0.01. The blue and red colors represent accumulation and depletion of hole density, respectively. Panels (a) and (b) are adapted from Refs.~\citenum{Kuleff2016} and~\citenum{Bruner2017}.}
    \label{fig:nitrosobenzene_charge_migration_comparison}
\end{figure}

\begin{figure}[!htbp]
    \centering
    \includegraphics[]{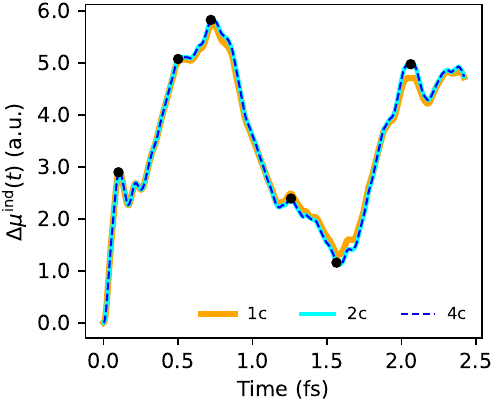}
    \caption{Nitrosobenzene: Time-dependent induced electric dipole moment (as in Eq.~\eqref{eq:dipole}\fixme{)}, following sudden ionization from N $1s$ orbital. The time-evolution is obtained at the non-relativistic 1c (orange), amfX2C 2c (cyan) and Dirac--Coulomb 4c (blue) Hamiltonian levels. The black points mark the time at which the hole density is plotted in Figure~\ref{fig:nitrosobenzene_charge_migration_comparison}. }
    \label{fig:nitrosobenzene_deltamu}
\end{figure}

Nitrosobenzene is a prototypical system in which charge migration following core-level ionization was first demonstrated theoretically by Kuleff \textit{et al.}~\cite{Kuleff2016} at timescales shorter than both Auger decay and the onset of nuclear-motion. The molecule has since served as a benchmark for subsequent theoretical developments in ultrafast electron dynamics, including RT-TDDFT.~\cite{Bruner2017} Following sudden ionization from the nitrogen K-edge, the molecule is prepared in a nonstationary cationic state that undergoes coherent electronic evolution.
As shown in Figure~\ref{fig:nitrosobenzene_charge_migration_comparison}, the initially localized hole density rapidly spreads over the adjacent $\pi$-conjugated benzene framework. Appreciable redistribution is already visible at $\sim \unit[0.12]{fs}$ and becomes pronounced at 0.48--\unit[0.70]{fs}. The subsequent snapshots show a reversal of this motion, followed by the recovery of a similar spatial distribution at \unit[2.06]{fs}. The hole density distributions at 0.70 and \unit[2.06]{fs} are quite similar, corresponding to a recurrence interval of $\sim\unit[1.36]{fs}$. This spatial pattern and its characteristic timescale are consistent with the earlier ADC(4) and PBE0/Sadlej+ calculations shown in Figure~\ref{fig:nitrosobenzene_charge_migration_comparison}a,b, which likewise show rapid delocalization of the initially localized core-hole perturbation into the aromatic $\pi$-system, followed by an oscillatory evolution of the hole density.
Even though the hole density provides a direct real-space representation of the electron dynamics, it is a three-dimensional spatial quantity, and becomes \fixme{quickly} difficult to analyze using individual snapshots alone. The problem is further exacerbated  when several delocalized orbitals are involved.

In order to address this, we
use the magnitude of the time-dependent induced electric dipole moment, $\vec{\mu}^{\mathrm{ind}}(t)$, as a compact scalar descriptor for charge migration. The changes in the electronic dipole moment originate directly from the redistribution of the electronic density and its magnitude is evaluated as
\begin{equation}
    \Delta{\mu}^{\text{ind}}(t)
    =
    \left| \Delta \vec{\mu}^{\text{ind}}(t)\right|
    =
    \Big| \text{Tr}\left[
        \mat{P}\big(\mat{D}_{+}^{\text{AO}}(t) - \mat{D}_{+}^{\text{AO}}(0)\big)
    \right] \Big|
    \label{eq:dipole}
\end{equation}
where $\mat{P}$ is the electric dipole moment matrix in the AO basis.
\fixme{Interestingly, a similar connection between dipole oscillations and charge migration was previously discussed for a longer femtosecond propagation window.\cite{Worner2017}. The induced dipole dynamics are also relevant to experimental observables such as attosecond transient absorption,\cite{Moitra2023} high-harmonic spectroscopy,\cite{Kraus2015} and time-resolved X-ray absorption.\cite{Moitra2023}}
This connection is illustrated in Figure~\ref{fig:nitrosobenzene_deltamu}, where the black markers denote the same time points at which the hole density is shown in Figure~\ref{fig:nitrosobenzene_charge_migration_comparison}. The rapid initial spreading of the hole density is accompanied by a sharp increase in $\Delta\mu^{\mathrm{ind}}(t)$, which reaches a pronounced maximum around \unit[0.7]{fs}. The subsequent reversal of the charge redistribution is reflected in a decrease of the induced dipole toward its minimum near \unit[1.5]{fs}, followed by a renewed increase as the density approaches a spatial distribution similar to that observed around \unit[0.7]{fs}. Thus, the prominent features of $\Delta\mu^{\text{ind}}(t)$ track the principal stages of the real-space charge migration, allowing the underlying electronic motion to be identified without relying solely on individual density snapshots.

We further observe that the induced electric dipole moments obtained at the 1c (orange), 2c (cyan), and 4c (blue) Hamiltonian levels remain nearly the same over the simulated time window, as shown in Figure~\ref{fig:nitrosobenzene_deltamu}.
This agreement provides a useful benchmark for the implementation, while also indicating that the charge migration in $K$-edge (1s) core-ionized nitrosobenzene is largely insensitive to the relativistic treatment over this timescale. This is consistent with the nature of the system and ionization induced dynamics. Nitrosobenzene is composed entirely of light elements, for which relativistic effects are small, and also the $1s$ core level is only affected by scalar relativistic shifts. As a result, the relativistic effects modify the underlying orbital energies, but they do not alter the coherent electron redistribution in this case.

\subsection{Iodoacetylene}\label{sec:iodoacetylene}
\begin{figure}[h]
    \centering
    \includegraphics[]{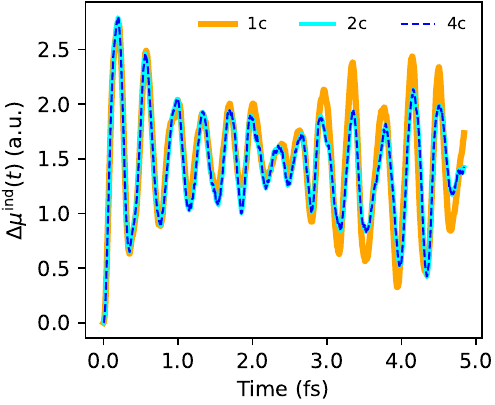}
    \caption{Iodoacetylene: Time-dependent induced electric dipole moment (as in Eq.~\eqref{eq:dipole}) following sudden ionization from the I $2p$ orbital. The time-evolution is obtained at the non-relativistic 1c (orange), amfX2C 2c (cyan) and Dirac--Coulomb 4c (blue) Hamiltonian levels. }
    \label{fig:iodoacetylene_mu}
\end{figure}
\begin{figure}[!htbp]
    \centering
    \includegraphics[width=\textwidth]{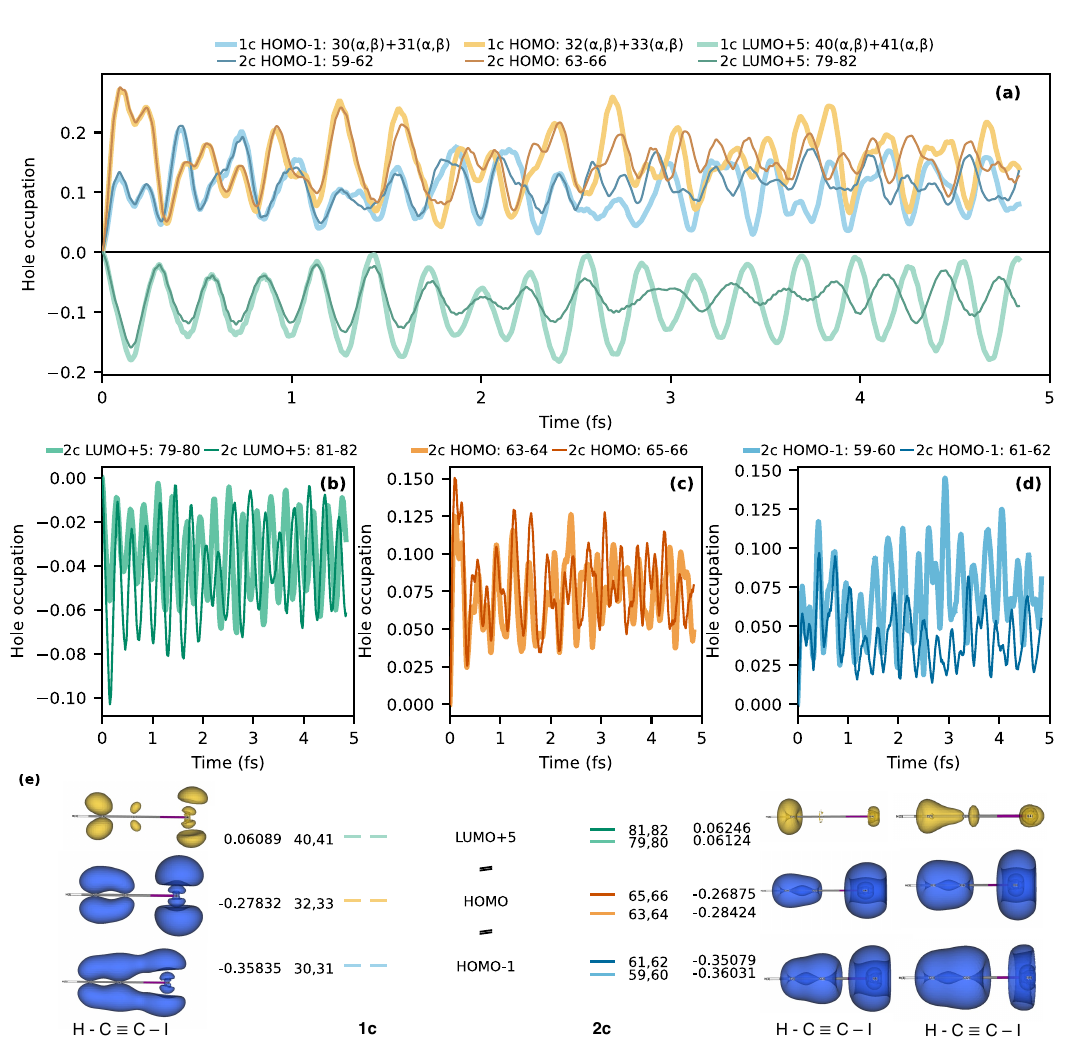}
    \caption{Iodoacetylene: Time-dependent molecular-orbital occupations of dominant contributors (obtained as Eq.~\eqref{eq:hole-occupation-MO}) governing the charge migration dynamics following I $2p$ electron removal.
(a) Time-dependent hole occupations obtained at the 1c and 2c levels for the HOMO$-1$, HOMO, and LUMO$+5$ manifolds, with the corresponding degenerate 1c orbitals and 2c spinors combined for direct comparison.
(b--d) Corresponding 2c dynamics resolved into the individual Kramers pairs arising from the relativistic splitting of the LUMO$+5$, HOMO, and HOMO$-1$ manifolds, respectively.
(e) 1c and 2c orbital-energy levels and orbital charge densities for the three dominant manifolds using an iso-value of 0.01. Energies are given in au.}
    \label{fig:iodoacetylene_mo}
\end{figure}

We next consider iodoacetylene, for which charge migration following valence ionization has been experimentally detected.\cite{Kraus2015} The presence of heavy-element iodine also makes it a \fixme{suitable} system for examining relativistic effects. The initial core hole state is prepared by removing electron from the I $2p$ orbital. The propagation is extended beyond the first fs to resolve the build-up of relativistic imprints during the coherent electronic evolution.
Auger-Meitner decay is not included in the present treatment and may become relevant on comparable timescales.
\fixme{Therefore, the longer-time dynamics
are interpreted as a proof-of-principle investigation
of the intrinsic coherent electronic response rather
than a prediction of experimentally observable
charge migration.}

Figure~\ref{fig:iodoacetylene_mu} compares the induced electric dipole moment dynamics obtained using 1c (orange), 2c (cyan), and 4c (blue) Hamiltonians. During approximately the first \unit[1.5]{fs}, the three methods give similar results. At later times, however, the non-relativistic 1c trajectory progressively deviates from the relativistic 2c and 4c ones. The differences appear mainly in the oscillation amplitudes, whereas the positions of the successive extrema remain comparable. The relativistic imprint becomes progressively more visible as the coherent evolution proceeds. The close agreement between the 2c and 4c results also gives confidence in the amfX2C treatment for describing the relativistic dynamics.

In order to understand the orbital origin of this relativistic signature, we compute the time-dependent variation of the ground-state molecular-orbital occupations, as shown in Figure~\ref{fig:iodoacetylene_mo}. The largest occupation changes involve the HOMO$-1$ (blue), HOMO (orange), and LUMO$+5$ (green) manifolds (Figure~\ref{fig:iodoacetylene_mo}a), all of which have $\pi$-character. At the 1c level, the HOMO$-1$ (orbitals 30 and 31), HOMO (32 and 33), and LUMO$+5$ (40 and 41) form degenerate manifolds. At the 2c level, each of these manifold splits into two Kramers pairs through spin--orbit (SO) splitting. The HOMO$-1$ gives Kramers pairs 59,60 and 61,62 at $-0.36031$ and -0.35079 au, respectively, while the HOMO gives 63,64 and 65,66 at -0.28424 and -0.26875 au. The LUMO$+5$ is more weakly SO-split into 79,80 and 81,82 at 0.06124 and 0.06246 au. In Figure~\ref{fig:iodoacetylene_mo}a, the contributions from the corresponding Kramers pairs are summed at the 2c level to allow a direct comparison with the degenerate 1c $\pi$-manifolds. The 1c and 2c occupation dynamics remain similar up to about \unit[1.5]{fs}, after which the differences become more apparent. The pair resolved dynamics in Figure~\ref{fig:iodoacetylene_mo}b--d show that the two components within each relativistically SO-split manifold follow distinct time-dependent evolutions.
This delayed separation can be understood from the phase evolution of the split states. Once a degenerate 1c manifold is split at the 2c level, its Kramers-pair components evolve with different energies and therefore accumulate a relative phase according to $\Delta\phi(t)=\Delta Et/\hbar.$
The phase difference is necessarily small at very short times and small energy separations. So, the summed 2c Kramers-pair occupations initially remain close to the corresponding 1c dynamics even though the underlying relativistic states are already energetically distinct. As the phase difference accumulates, the interference between the participating orbital contributions changes, leading to bigger differences in their occupations and hence the induced dipole dynamics.

This factor is unexplored in the literature studies involving core-hole induced charge migration dynamics. Non-relativistic treatments restricted to the first femtosecond can capture the initial electronic response accurately, but that does not establish the absence of relativistic imprint. Rather, the observable relativistic effect develops as phase differences between the relativistically split valence components accumulate during the propagation, and often has a delayed emergence.

\subsection{Interhalogen compounds} \label{sec:interhalogen}

\begin{figure}[!htbp]
    \centering
    \includegraphics[width=\textwidth]{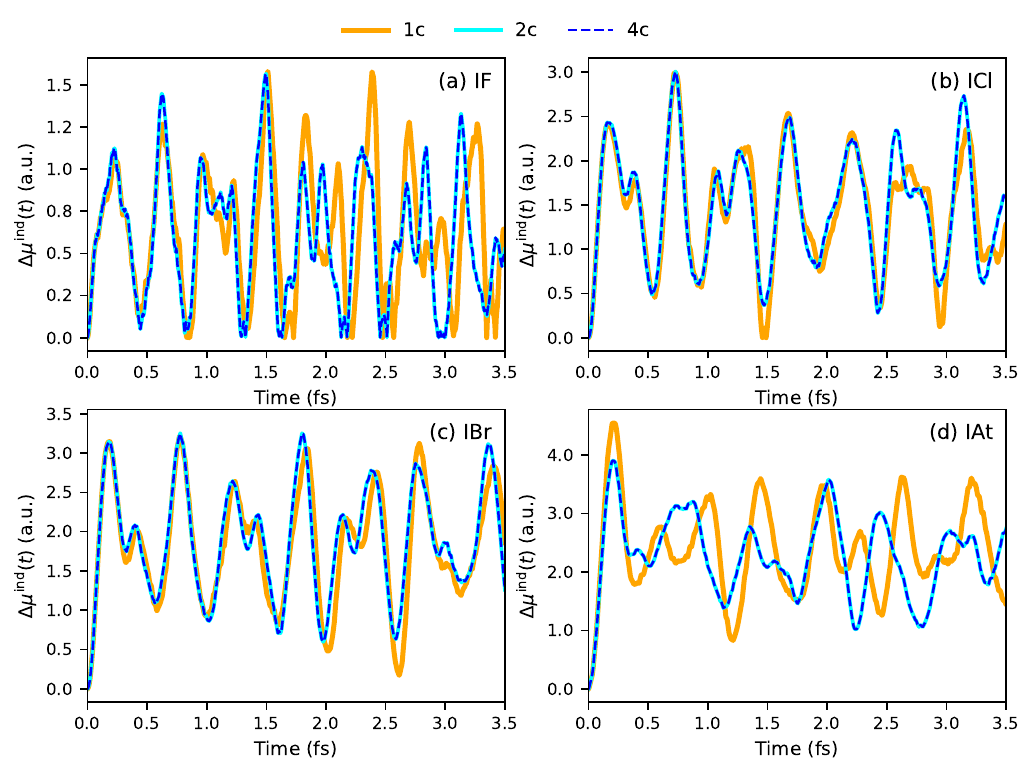}
    \caption{Iodohalides: Time-dependent induced electric dipole moment  of the cationic non-stationary state (obtained as Eq.~\eqref{eq:dipole}) created by electron removal from I $2p$ orbital for (a) IF (b) ICl (c) IBr and (d) IAt. The dynamics are obtained using the non-relativistic 1c (orange), amfX2c 2c (cyan) and Dirac--Coulomb 4c (blue) Hamiltonians. }
    \label{fig:iodohalides}
\end{figure}

\begin{figure}[!htbp]
    \centering
    \includegraphics[width=\textwidth]{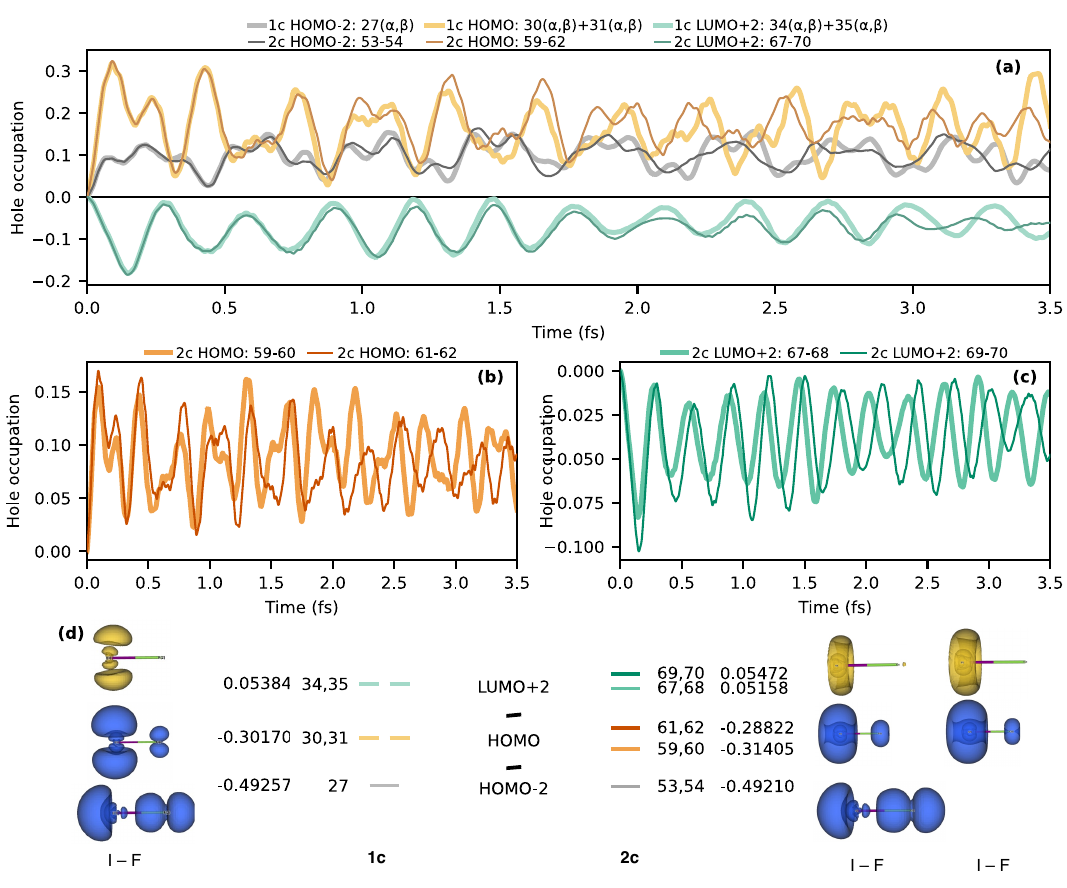}
    \caption{IF: Time-dependent molecular-orbital occupations of dominant contributors  (obtained as Eq.~\eqref{eq:hole-occupation-MO}) governing the charge migration dynamics following I $2p$ electron removal.
(a) Time-dependent hole occupations obtained at the 1c and 2c levels for the HOMO$-2$ (gray), HOMO (yellow), and LUMO$+2$ (green) manifolds, with the corresponding degenerate 1c orbitals and 2c spinors combined to preserve the electron count.
(b--c) Corresponding 2c dynamics resolved into individual Kramers pairs arising from the relativistic splitting of the HOMO and LUMO$+2$ manifolds, respectively.
(d) Corresponding 1c and 2c orbital-energy levels and orbital charge densities for the three dominant manifolds using an iso-value of 0.01. Energies are given in au.}
    \label{fig:IF_mos}
\end{figure}

We next study the dynamic evolution of the cationic nonstationary state created by I $2p$ electron removal in the interhalogen series IF, ICl, IBr and IAt.
Figure~\ref{fig:iodohalides} compares the time-dependent induced electric dipole moments obtained using the one-component (1c), two-component (2c) and four-component (4c) Hamiltonians. Across the series, the 2c and 4c trajectories remain quantitatively similar.

For IF, the 1c and 2c results remain similar during approximately the first \unit[1.0]{fs}. At later times, however, clear differences appear in both the amplitudes and temporal positions of the extrema. A time-dependent molecular orbital contribution analysis shows that the dominant occupation changes involve HOMO-2 (orbital 27, gray) , HOMO (orbitals 30 and 31, yellow) and LUMO+2 (orbitals 34 and 35, green) manifolds, shown in Figure~\ref{fig:IF_mos}a. The degenerate HOMO and LUMO+2 pairs are split at the 2c level, with energy splitting of 0.7 eV and 0.085 eV, respectively.
The spin-orbit split 2c orbital dynamics are presented in Figure~\ref{fig:IF_mos}b,c.
The relativistic splitting is manifested dynamically through different amplitudes and phases of the corresponding Kramers-pair occupations, which become increasingly distinct during propagation.

\begin{figure}
    \centering
    \includegraphics[width=\textwidth]{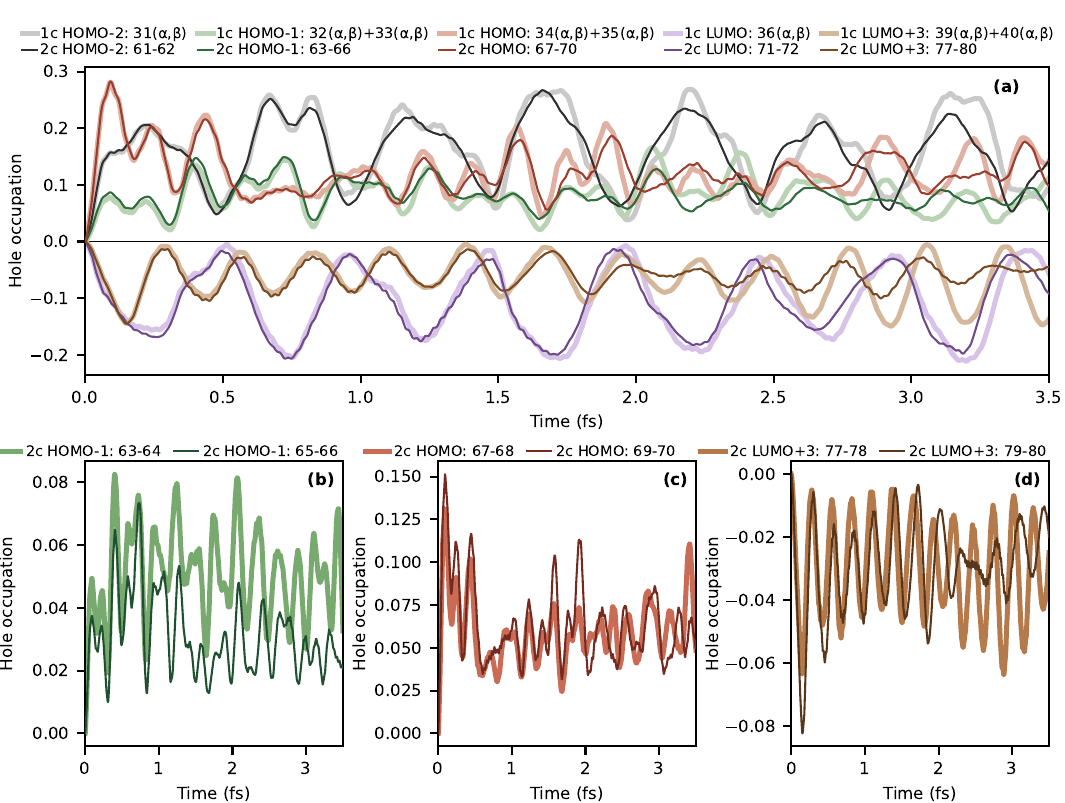}
    \caption{ICl: Time-dependent molecular-orbital occupations of dominant contributors (obtained as Eq.~\eqref{eq:hole-occupation-MO}) governing the charge migration dynamics following I $2p$ electron removal.
(a) Time-dependent hole occupations obtained at the 1c and 2c levels for the HOMO$-2$ (gray), HOMO-1 (green), HOMO (maroon), LUMO (purple) and LUMO+3 (brown) manifolds, with the corresponding degenerate 1c orbitals and 2c spinors combined to preserve the electron count.
(b--d) Corresponding 2c dynamics resolved into individual Kramers pairs arising from the relativistic splitting of the HOMO-1, HOMO and LUMO+3 manifolds, respectively.
}
    \label{fig:ICl_I2p}
\end{figure}

Contrarily, the relativistic imprint is substantially weaker for ICl and IBr, as shown in Figure~\ref{fig:iodohalides}b,c. A molecular orbital analysis for ICl is shown in Figure~\ref{fig:ICl_I2p}, shows that the dynamics is controlled by non-degenerate HOMO-2 and LUMO orbitals. The contributions from the degenerate $\pi$ manifolds still show relativistic splittings (Figure~\ref{fig:ICl_I2p}b,c,d), \fixme{even though their} contribution to the overall dynamics is weak.
The strongest and most persistent relativistic signature is observed for IAt. Although the initial rise of the induced dipole moment is similar at the 1c and 2c/4c levels up to \unit[0.4]{fs}, the trajectories separate rapidly afterwards. Considerable differences subsequently occur in the amplitudes and temporal positions of the oscillations.
The variation across the series is clearly non-monotonic. This shows that the magnitude of the relativistic effect cannot be predicted solely from the presence of heavy element, but depends on the valence molecular orbital manifold participating in the charge re-distribution.  The molecular orbital occupation analysis for IBr and IAt are presented in the Supporting information.

\begin{figure}[!htbp]
    \centering
    \includegraphics[width=\textwidth]{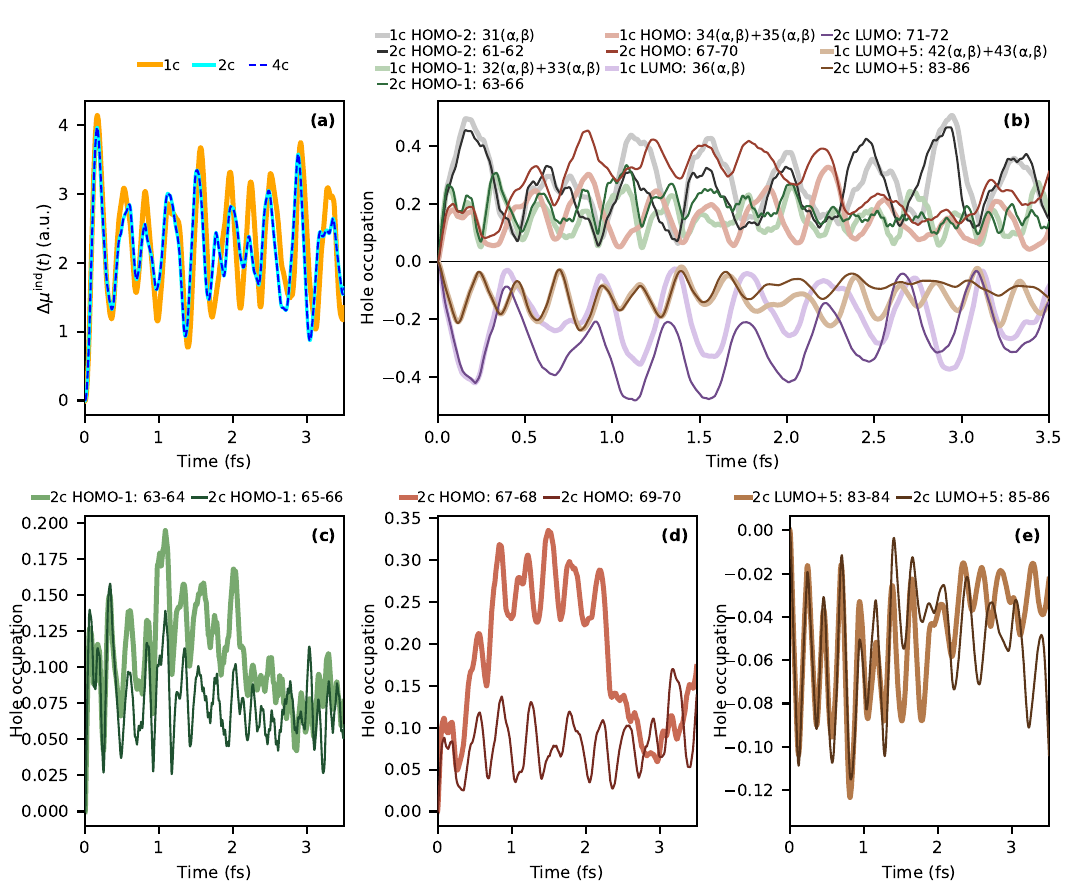}
    \caption{ICl: (a) Time-dependent induced electric dipole moment (obtained as Eq.~\eqref{eq:dipole}) of the cationic non-stationary state created by electron removal from Cl $2p$ orbital.
    (b) Time-dependent molecular orbital occupations (obtained as Eq.~\eqref{eq:hole-occupation-MO}) obtained at the 1c and 2c levels for the dominant contributors, with the corresponding degenerate 1c orbitals and 2c spinors combined for direct comparison.
(c--e) Corresponding 2c dynamics resolved into the individual Kramers pairs arising from the relativistic splitting of the HOMO-1, HOMO and LUMO+5 manifolds, respectively.
    }
    \label{fig:ICl_Cl2p}
\end{figure}

Additionally, we investigated charge migration initiated by ionization of the Cl $2p$ shell in ICl. This provides a complementary core-ionization channel. The Cl $2p$ core hole has a longer lifetime than the I $2p$ core hole, which provides a wider experimental window for observing the coherent electron dynamics. The core-hole decay itself is not included in the theoretical treatment.
The corresponding results are presented in Figure~\ref{fig:ICl_Cl2p}. In contrast to I $2p$ ionization, the Cl $2p$ ionization produces a noticeable difference between the non-relativistic 1c and relativistic 2c and 4c results after approximately \unit[1.5]{fs} (Figure~\ref{fig:ICl_Cl2p}a). The relativistic imprint is therefore strongly dependent on the initial ionization channel, even within the same molecular system.

The molecular-orbital-resolved analysis in Figure~\ref{fig:ICl_Cl2p}b identifies the HOMO$-2$ (orbital 31, gray), HOMO$-1$ (orbitals 32 and 33, green), HOMO (orbitals 34 and 35, maroon), LUMO (orbital 36, purple), and LUMO$+5$ (orbitals 42 and 43, brown) as the principal contributors to the charge redistribution.
Of these, the HOMO-2 and LUMO are the major contributors and are non-degenerate orbitals.
The pair-resolved dynamics in Figure~\ref{fig:ICl_Cl2p}c,d,e consequently show markedly different time-dependent occupations for the two non-degenerate Kramers pair levels of the split 1c degenerate states. This concertedly contributes to the difference in charge migration timescales beyond the initial \unit[1.5]{fs}.

\section{Conclusions}
In summary, we have presented the theory and implementation of a relativistic RT-TDDFT framework for core-ionization induced charge migration dynamics and used it to investigate how relativistic effects manifest in this process. We found that the induced electric dipole provides a compact spectroscopically relevant measure of the dominant charge redistribution, capable of reproducing the characteristic timescales for charge migration. The implementation was benchmarked using the nitrosobenzene molecule. Relativistic effects were then examined for I $2p$-ionized iodoacetylene and the interhalogen series, together with the complementary Cl $2p$-ionization channel in ICl. The influence of relativity was found to be strongly dependent on the system and ionization channel. Pronounced differences between relativistic and nonrelativistic dynamics are observed for IF and IAt, whereas ICl and IBr remain insensitive to the relativistic treatment following I $2p$ ionization. In contrast, Cl $2p$ ionization of ICl gives rise to strong relativistic imprints. The molecular orbital resolved analysis shows that these differences arise from relativistic spin-orbit splitting of the active valence orbital manifold. The agreement between the amfX2C (2c) and four-component (4c) dynamics throughout establishes amfX2C as an accurate and computationally cheaper approach for relativistic real-time electron dynamics.

An important feature observed in these time-evolutions is the delayed emergence of the relativistic effect, usually after 0.5--\unit[2]{fs}.
\fixme{These results provide a proof-of-principle demonstration}
that relativistic effects can leave an imprint on attosecond charge migration even when they are not immediately apparent in the earliest times. Future developments incorporating finite core-hole lifetimes, particularly Auger-Meitner decay, as well as nuclear motion will be important for establishing how these coherent relativistic signatures survive under experimentally relevant conditions.

\begin{acknowledgement}
T.M. acknowledges support from the Marie Skłodowska-Curie Individual Postdoctoral Fellowship (grant No.~101152113). We also acknowledge support from the Research Council of Norway through its Centres of Excellence scheme (project No.~262695) and research grant No.~315822. M.R. acknowledges funding from the EU NextGenerationEU through the Recovery and Resilience Plan for Slovakia under project No.~09I05-03-V02-00034, as well as from the Slovak Research and Development Agency (grant No.~APVV-25-0750) and VEGA (grant No.~1/0670/24). The computational resources used in this work were funded by the EuroHPC Regular Access grant (No.~EU-25-8) and by NextGenerationEU through the Recovery and Resilience Plan for Slovakia under project No.~17I03-04-P02-00001.
\end{acknowledgement}

\begin{suppinfo}
Time-dependent variation in molecular orbital analysis; geometries.
\end{suppinfo}

\FloatBarrier
\bibliography{article}

\end{document}